\documentclass[11pt,oneside,english,reqno]{amsart}
\usepackage[T1]{fontenc}
\usepackage{pdfsync}
\usepackage{verbatim}
\usepackage{amsthm}
\usepackage{amstext}
\usepackage{amssymb}
\usepackage{undertilde}
\usepackage{esint}
\usepackage{xargs}[2008/03/08]

\makeatletter
\numberwithin{equation}{section}

\newcommand{\ie}{\textit{i.e.}}

\usepackage{mathpazo}
\usepackage[cal=boondoxo,bb=boondox,frak=boondox]{mathalfa}%%
\makeatother

\usepackage{babel}
\begin{document}

\begin{comment}
Basic Math
\end{comment}

\global\long\def\ga{\alpha}
\global\long\def\gb{\beta}
\global\long\def\ggm{\gamma}
\global\long\def\go{\omega}
\global\long\def\ge{\epsilon}
\global\long\def\gs{\sigma}
\global\long\def\gd{\delta}
\global\long\def\gD{\Delta}
\global\long\def\vph{\varphi}
\global\long\def\gf{\varphi}
\global\long\def\gk{\kappa}
\global\long\def\eps{\varepsilon}
\global\long\def\epss#1#2{\varepsilon_{#2}^{#1}}
\global\long\def\ep#1{\eps_{#1}}
\global\long\def\wh#1{\widehat{#1}}
\global\long\def\spec#1{\textsf{#1}}
\global\long\def\ui{\wh{\boldsymbol{\imath}}}
\global\long\def\uj{\wh{\boldsymbol{\jmath}}}
\global\long\def\uk{\widehat{\boldsymbol{k}}}
\global\long\def\uI{\widehat{\mathbf{I}}}
\global\long\def\uJ{\widehat{\mathbf{J}}}
\global\long\def\uK{\widehat{\mathbf{K}}}
\global\long\def\bs#1{\boldsymbol{#1}}
\global\long\def\vect#1{\mathbf{#1}}
\global\long\def\bi#1{\textbf{\emph{#1}}}
\global\long\def\uv#1{\widehat{\boldsymbol{#1}}}
\global\long\def\cross{\times}
\global\long\def\ddt{\frac{\dee}{\dee t}}
\global\long\def\dbyd#1{\frac{\dee}{\dee#1}}
\global\long\def\dby#1#2{\frac{\partial#1}{\partial#2}}
\global\long\def\vct#1{\mathbf{#1}}
\begin{comment}
General Math
\end{comment}
\global\long\def\partialby#1#2{\frac{\partial#1}{\partial x^{#2}}}
\newcommandx\parder[2][usedefault, addprefix=\global, 1=]{\frac{\partial#2}{\partial#1}}
\global\long\def\oneto{1,\dots,}
\global\long\def\mi#1{\boldsymbol{#1}}
\global\long\def\mii{\mi I}
\begin{comment}
Multi-index
\end{comment}
\global\long\def\fall{,\quad\text{for all}\quad}
\global\long\def\reals{\mathbb{R}}
\global\long\def\rthree{\reals^{3}}
\global\long\def\rsix{\reals^{6}}
\global\long\def\rn{\reals^{n}}
\global\long\def\rt#1{\reals^{#1}}
\global\long\def\les{\leqslant}
\global\long\def\ges{\geqslant}
\global\long\def\dee{\textrm{d}}
\global\long\def\di{d}
\global\long\def\from{\colon}
\global\long\def\tto{\longrightarrow}
\global\long\def\abs#1{\left|#1\right|}
\global\long\def\isom{\cong}
\global\long\def\comp{\circ}
\global\long\def\cl#1{\overline{#1}}
\global\long\def\fun{\varphi}
\global\long\def\interior{\textrm{Int}\,}
\global\long\def\sign{\textrm{sign}\,}
\global\long\def\sgn#1{(-1)^{#1}}
\global\long\def\sgnp#1{(-1){}^{\abs{#1}}}
\global\long\def\dimension{\textrm{dim}\,}
\global\long\def\esssup{\textrm{ess}\,\sup}
\global\long\def\ess{\textrm{{ess}}}
\global\long\def\kernel{\mathop{\textrm{Kernel}}}
\global\long\def\support{\textrm{supp}\,}
\global\long\def\image{\textrm{Image}\,}
\global\long\def\diver{\mathop{\textrm{div}}}
\global\long\def\sp{\mathop{\textrm{span}}}
\global\long\def\resto#1{|_{#1}}
\global\long\def\incl{\iota}
\global\long\def\iden{\imath}
\global\long\def\idnt{\textrm{Id}}
\global\long\def\rest{\rho}
\global\long\def\extnd{e_{0}}
\global\long\def\proj{\textrm{pr}}
\global\long\def\ino#1{\int_{#1}}
\global\long\def\half{\frac{1}{2}}
\global\long\def\shalf{{\scriptstyle \half}}
\global\long\def\third{\frac{1}{3}}
\global\long\def\empt{\varnothing}
\global\long\def\paren#1{\left(#1\right)}
\global\long\def\bigp#1{\bigl(#1\bigr)}
\global\long\def\biggp#1{\biggl(#1\biggr)}
\global\long\def\Bigp#1{\Bigl(#1\Bigr)}
\global\long\def\braces#1{\left\{  #1\right\}  }
\global\long\def\sqbr#1{\left[#1\right]}
\global\long\def\anglep#1{\left\langle #1\right\rangle }
\global\long\def\lsum{{\textstyle \sum}}
\global\long\def\bigabs#1{\bigl|#1\bigr|}
\global\long\def\lisub#1#2#3{#1_{1}#2\dots#2#1_{#3}}
\global\long\def\lisup#1#2#3{#1^{1}#2\dots#2#1^{#3}}
\global\long\def\lisubb#1#2#3#4{#1_{#2}#3\dots#3#1_{#4}}
\global\long\def\lisubbc#1#2#3#4{#1_{#2}#3\cdots#3#1_{#4}}
\global\long\def\lisubbwout#1#2#3#4#5{#1_{#2}#3\dots#3\widehat{#1}_{#5}#3\dots#3#1_{#4}}
\global\long\def\lisubc#1#2#3{#1_{1}#2\cdots#2#1_{#3}}
\global\long\def\lisupc#1#2#3{#1^{1}#2\cdots#2#1^{#3}}
\global\long\def\lisupp#1#2#3#4{#1^{#2}#3\dots#3#1^{#4}}
\global\long\def\lisuppc#1#2#3#4{#1^{#2}#3\cdots#3#1^{#4}}
\global\long\def\lisuppwout#1#2#3#4#5#6{#1^{#2}#3#4#3\wh{#1^{#6}}#3#4#3#1^{#5}}
\global\long\def\lisubbwout#1#2#3#4#5#6{#1_{#2}#3#4#3\wh{#1}_{#6}#3#4#3#1_{#5}}
\global\long\def\lisubwout#1#2#3#4{#1_{1}#2\dots#2\widehat{#1}_{#4}#2\dots#2#1_{#3}}
\global\long\def\lisupwout#1#2#3#4{#1^{1}#2\dots#2\widehat{#1^{#4}}#2\dots#2#1^{#3}}
\global\long\def\lisubwoutc#1#2#3#4{#1_{1}#2\cdots#2\widehat{#1}_{#4}#2\cdots#2#1_{#3}}
\global\long\def\twp#1#2#3{\dee#1^{#2}\wedge\dee#1^{#3}}
\global\long\def\thp#1#2#3#4{\dee#1^{#2}\wedge\dee#1^{#3}\wedge\dee#1^{#4}}
\global\long\def\fop#1#2#3#4#5{\dee#1^{#2}\wedge\dee#1^{#3}\wedge\dee#1^{#4}\wedge\dee#1^{#5}}
\global\long\def\idots#1{#1\dots#1}
\global\long\def\icdots#1{#1\cdots#1}
\global\long\def\norm#1{\|#1\|}
\global\long\def\nonh{\heartsuit}
\global\long\def\nhn#1{\norm{#1}^{\nonh}}
\global\long\def\trps{^{{\scriptscriptstyle \textsf{T}}}}
\global\long\def\testfuns{\mathcal{D}}
\global\long\def\ntil#1{\tilde{#1}{}}
\begin{comment}
Forms-Differential Geometry
\end{comment}
\global\long\def\alt{\mathfrak{A}}
\global\long\def\pou{\eta}
\global\long\def\ext{{\textstyle \bigwedge}}
\global\long\def\forms{\Omega}
\global\long\def\dotwedge{\dot{\mbox{\ensuremath{\wedge}}}}
\global\long\def\vel{\theta}
\begin{comment}
<volume element
\end{comment}
\global\long\def\contr{\raisebox{0.4pt}{\mbox{\ensuremath{\lrcorner}}}\,}
\global\long\def\fcontr{\raisebox{0.4pt}{\mbox{\ensuremath{\llcorner}}}\,}
\global\long\def\lie{\mathcal{L}}
\global\long\def\L#1{L\bigl(#1\bigr)}
\global\long\def\vvforms{\ext^{\dims}\bigp{T\spc,\vbts^{*}}}
\begin{comment}
>\textcompwordmark{}>\textcompwordmark{}>\textcompwordmark{}>\textcompwordmark{}>\textcompwordmark{}>Space
Time Events<\textcompwordmark{}<\textcompwordmark{}<\textcompwordmark{}<\textcompwordmark{}<\textcompwordmark{}<\textcompwordmark{}<\textcompwordmark{}<
\end{comment}
\global\long\def\spc{\mathcal{S}}
\global\long\def\sptm{\mathcal{E}}
\global\long\def\evnt{e}
\global\long\def\frame{\Phi}
\global\long\def\timeman{\mathcal{T}}
\global\long\def\zman{t}
\global\long\def\dims{d}
\global\long\def\m{\dims-1}
\global\long\def\dimw{m}
\global\long\def\wc{z}
\global\long\def\fourv#1{\mbox{\ensuremath{\mathfrak{#1}}}}
\global\long\def\pbform#1{\utilde{#1}}
\global\long\def\util#1{\raisebox{-5pt}{\ensuremath{{\scriptscriptstyle \sim}}}\!\!\!#1}
\global\long\def\utilJ{\util J}
\global\long\def\utilRho{\util{\rho}}
\global\long\def\body{B}
\global\long\def\man{\mathcal{M}}
\global\long\def\var{\mathcal{V}}
\global\long\def\bdry{\partial}
\global\long\def\gO{\varOmega}
\global\long\def\reg{\mathcal{R}}
\global\long\def\bdrr{\bdry\reg}
\global\long\def\bdom{\bdry\gO}
\global\long\def\bndo{\partial\gO}
\begin{comment}
{*}{*}{*}{*}{*}{*}{*}{*}{*}{*}{*}{*}{*}{*}{*}{*}{*}{*}{*}{*}{*}{*}{*}{*}{*}{*}{*}{*}{*}{*}{*}{*}{*}{*}{*}{*}{*}{*}{*}{*}{*}{*}
{*}{*}{*}{*}{*}{*}{*}{*}{*}{*}{*}{*}{*}{*}{*}{*}{*}{*}{*}Cauchy Fluxes{*}{*}{*}{*}{*}{*}{*}{*}{*}{*}
{*}{*}{*}{*}{*}{*}{*}{*}{*}{*}{*}{*}{*}{*}{*}{*}{*}{*}{*}{*}{*}{*}{*}{*}{*}{*}{*}{*}{*}{*}{*}{*}{*}{*}{*}{*}{*}{*}{*}{*}{*}{*}{*}
\end{comment}
\global\long\def\pform{\varsigma}
\global\long\def\vform{\beta}
\global\long\def\sform{\tau}
\global\long\def\flow{J}
\global\long\def\n{\m}
\global\long\def\cmap{\mathfrak{t}}
\global\long\def\vcmap{\varSigma}
\global\long\def\mvec{\mathfrak{v}}
\global\long\def\mveco#1{\mathfrak{#1}}
\global\long\def\smbase{\mathfrak{e}}
\global\long\def\spx{\simp}
\global\long\def\hp{H}
\global\long\def\ohp{h}
\global\long\def\hps{G_{\dims-1}(T\spc)}
\global\long\def\ohps{G_{\dims-1}^{\perp}(T\spc)}
\global\long\def\hpsx{G_{\dims-1}(\tspc)}
\global\long\def\ohpsx{G_{\dims-1}^{\perp}(\tspc)}
\global\long\def\fbun{F}
\global\long\def\flowm{\Phi}
\global\long\def\tgb{T\spc}
\global\long\def\ctgb{T^{*}\spc}
\global\long\def\tspc{T_{\pis}\spc}
\global\long\def\dspc{T_{\pis}^{*}\spc}
\begin{comment}
{*}{*}{*}{*}{*} ELECTROMAGNETISM IN SPACETIME <\textcompwordmark{}<\textcompwordmark{}<\textcompwordmark{}<\textcompwordmark{}<\textcompwordmark{}<
\end{comment}
%%%%%%% ELECTROMAGNETISM IN SPACETIME %%%%%%
\global\long\def\fflow{\fourv J}
%      four-flow
\global\long\def\fvform{\mathfrak{b}}
%      four body flux
\global\long\def\fsform{\mathfrak{t}}
%      four surface flux
\global\long\def\fpform{\mathfrak{s}}
%      Four production rate
%\newcommand{\fgr}{\mathfrak{I}}%         Growth rate in space time
\global\long\def\maxw{\mathfrak{g}}
%        Maxwell form or Stream form
\global\long\def\frdy{\mathfrak{f}}
%        Faraday 2-form
%\newcommand{\maxw}{\mathcal{G}}%        Maxwell form or Stream form
%\newcommand{\frdy}{\mathcal{F}}%        Faraday 2-form
\global\long\def\ptnl{A}
%                   Potential 1-form
%%%%%%%%%%%%%%%%%%%%%%%%%%%%%%%%%%%%%%%%%%%%%
\begin{comment}
{*}{*}{*}{*}{*} Jets and Vector Bundles {*}{*}{*}{*}{*}
\end{comment}
\global\long\def\eucl{E}
\global\long\def\mind{\alpha}
\global\long\def\vb{\xi}
\global\long\def\man{\mathcal{M}}
\global\long\def\odman{\mathcal{N}}
\global\long\def\subman{\mathcal{A}}
\global\long\def\vbt{\mathcal{E}}
\global\long\def\fib{\mathbf{V}}
\global\long\def\vbts{W}
\global\long\def\avb{U}
\global\long\def\chart{\varphi}
\global\long\def\vbchart{\Phi}
\global\long\def\jetb#1{J^{#1}}
\global\long\def\jet#1{j^{1}(#1)}
\global\long\def\Jet#1{J^{1}(#1)}
\global\long\def\jetm#1{j_{#1}}
\begin{comment}
Sobolev Spaces
\end{comment}
\global\long\def\sobp#1#2{W_{#2}^{#1}}
\global\long\def\inner#1#2{\left\langle #1,#2\right\rangle }
\global\long\def\fields{\sobp pk(\vb)}
\global\long\def\bodyfields{\sobp p{k_{\partial}}(\vb)}
\global\long\def\forces{\sobp pk(\vb)^{*}}
\global\long\def\bfields{\sobp p{k_{\partial}}(\vb\resto{\bndo})}
\global\long\def\loadp{(\sfc,\bfc)}
\global\long\def\strains{\lp p(\jetb k(\vb))}
\global\long\def\stresses{\lp{p'}(\jetb k(\vb)^{*})}
\global\long\def\diffop{D}
\global\long\def\strainm{E}
\global\long\def\incomps{\vbts_{\yieldf}}
\global\long\def\devs{L^{p'}(\eta_{1}^{*})}
\global\long\def\incompsns{L^{p}(\eta_{1})}
\begin{comment}
Distributions and Currents
\end{comment}
\global\long\def\testf{\mathcal{D}}
\global\long\def\dists{\mathcal{D}'}
\global\long\def\codiv{\boldsymbol{\partial}}
\global\long\def\currof#1{\tilde{#1}}
\global\long\def\chn{c}
\global\long\def\chnsp{\mathbf{F}}
\global\long\def\current{T}
\global\long\def\curr#1{T_{\langle#1\rangle}}
\global\long\def\prop{P}
\global\long\def\aprop{Q}
\global\long\def\flux{T}
\global\long\def\aflux{S}
\global\long\def\fform{\tau}
\global\long\def\dimn{n}
\global\long\def\sdim{{\dimn-1}}
\global\long\def\contrf{{\scriptstyle \smallfrown}}
\global\long\def\prodf{{\scriptstyle \smallsmile}}
\global\long\def\ptnl{\varphi}
\global\long\def\form{\omega}
\global\long\def\dens{\rho}
\global\long\def\simp{s}
\global\long\def\ssimp{\Delta}
\global\long\def\cpx{K}
\global\long\def\cell{C}
\global\long\def\chain{B}
\global\long\def\ach{A}
\global\long\def\coch{X}
\global\long\def\scale{s}
\global\long\def\fnorm#1{\norm{#1}^{\flat}}
\global\long\def\chains{\mathcal{A}}
\global\long\def\ivs{\boldsymbol{U}}
\global\long\def\mvs{\boldsymbol{V}}
\global\long\def\cvs{\boldsymbol{W}}
\begin{comment}
Points Vectors and Regions
\end{comment}
\global\long\def\pis{x}
\global\long\def\xo{\pis_{0}}
\global\long\def\pib{X}
\global\long\def\pbndo{\Gamma}
\global\long\def\bndoo{\pbndo_{0}}
 \global\long\def\bndot{\pbndo_{t}}
\global\long\def\cloo{\cl{\gO}}
\global\long\def\nor{\mathbf{n}}
\global\long\def\dA{\,\dee A}
\global\long\def\dV{\,\dee V}
\global\long\def\eps{\varepsilon}
\global\long\def\vs{\mathbf{W}}
\global\long\def\avs{\mathbf{V}}
\global\long\def\affsp{\mathbf{A}}
\global\long\def\pt{p}
\global\long\def\vbase{e}
\global\long\def\sbase{\mathbf{e}}
\global\long\def\msbase{\mathfrak{e}}
\global\long\def\vect{v}
\begin{comment}
Kinematics, Strains
\end{comment}
\global\long\def\vf{w}
\global\long\def\avf{u}
\global\long\def\stn{\varepsilon}
\global\long\def\rig{r}
\global\long\def\rigs{\mathcal{R}}
\global\long\def\qrigs{\!/\!\rigs}
\global\long\def\qd{\!/\,\!\kernel\diffop}
\global\long\def\dis{\chi}
\global\long\def\conf{\kappa}
\begin{comment}
Forces and Stresses
\end{comment}
\global\long\def\fc{F}
\global\long\def\st{\sigma}
\global\long\def\bfc{\mathbf{b}}
\global\long\def\sfc{\mathbf{t}}
\global\long\def\stm{S}
\global\long\def\nhs{Y}
\begin{comment}
Nonholonomic Variational Second Order Stress
\end{comment}
\global\long\def\soc{Z}
\begin{comment}
Second Order Cauchy Stress
\end{comment}
\global\long\def\tran{\mathrm{tr}}
\global\long\def\slf{R}
%Self force
\global\long\def\sts{\varSigma}
%stresses
\global\long\def\ebdfc{T}
\global\long\def\optimum{\st^{\textrm{opt}}}
\global\long\def\scf{K}
\begin{comment}
Function Spaces
\end{comment}
\global\long\def\cee#1{C^{#1}}
\global\long\def\lone{L^{1}}
\global\long\def\linf{L^{\infty}}
\global\long\def\lp#1{L^{#1}}
\global\long\def\ofbdo{(\bndo)}
\global\long\def\ofclo{(\cloo)}
\global\long\def\vono{(\gO,\rthree)}
\global\long\def\vonbdo{(\bndo,\rthree)}
\global\long\def\vonbdoo{(\bndoo,\rthree)}
\global\long\def\vonbdot{(\bndot,\rthree)}
\global\long\def\vonclo{(\cl{\gO},\rthree)}
\global\long\def\strono{(\gO,\reals^{6})}
\global\long\def\sob{W_{1}^{1}}
\global\long\def\sobb{\sob(\gO,\rthree)}
\global\long\def\lob{\lone(\gO,\rthree)}
\global\long\def\lib{\linf(\gO,\reals^{12})}
\global\long\def\ofO{(\gO)}
\global\long\def\oneo{{1,\gO}}
\global\long\def\onebdo{{1,\bndo}}
\global\long\def\info{{\infty,\gO}}
\global\long\def\infclo{{\infty,\cloo}}
\global\long\def\infbdo{{\infty,\bndo}}
\global\long\def\ld{LD}
\global\long\def\ldo{\ld\ofO}
\global\long\def\ldoo{\ldo_{0}}
\global\long\def\trace{\gamma}
\global\long\def\pr{\proj_{\rigs}}
\global\long\def\pq{\proj}
\global\long\def\qr{\,/\,\reals}
\begin{comment}
Plasticity and Optimization
\end{comment}
\global\long\def\aro{S_{1}}
\global\long\def\art{S_{2}}
\global\long\def\mo{m_{1}}
\global\long\def\mt{m_{2}}
\begin{comment}
Optimization
\end{comment}
\global\long\def\yieldc{B}
\global\long\def\yieldf{Y}
\global\long\def\trpr{\pi_{P}}
\global\long\def\devpr{\pi_{\devsp}}
\global\long\def\prsp{P}
\global\long\def\devsp{D}
\global\long\def\ynorm#1{\|#1\|_{\yieldf}}
\global\long\def\colls{\Psi}
%Collapse sufrace
\begin{comment}
Finite Elements
\end{comment}
\global\long\def\ssx{S}
\global\long\def\smap{s}
\global\long\def\smat{\chi}
\global\long\def\sx{e}
\global\long\def\snode{P}
\global\long\def\elem{e}
\global\long\def\nel{L}
\global\long\def\el{l}
\global\long\def\ipln{\phi}
\global\long\def\ndof{D}
\global\long\def\dof{d}
\global\long\def\nldof{N}
\global\long\def\ldof{n}
\global\long\def\lvf{\chi}
\global\long\def\lfc{\varphi}
\global\long\def\amat{A}
\global\long\def\snomat{E}
\global\long\def\femat{E}
\global\long\def\tmat{T}
\global\long\def\fvec{f}
\global\long\def\snsp{\mathcal{S}}
\global\long\def\slnsp{\Phi}
\global\long\def\ro{r_{1}}
\global\long\def\rtwo{r_{2}}
\global\long\def\rth{r_{3}}
\global\long\def\subbs{\mathcal{B}}
\global\long\def\elements{\mathcal{E}}
\global\long\def\element{E}
\global\long\def\nodes{\mathcal{N}}
\global\long\def\node{N}
\global\long\def\psubbs{\mathcal{P}}
\global\long\def\psubb{P}
\global\long\def\matr{M}
\global\long\def\nodemap{\nu}
\begin{comment}
{*}{*}{*}{*}{*}{*}
FINITE CHAINS
{*}{*}{*}{*}{*}{*}{*}{*}
\end{comment}
\global\long\def\node{v}
\global\long\def\edge{e}
\global\long\def\accu{q}
\global\long\def\accusp{\mathcal{Q}}
\global\long\def\potl{\varphi}
\global\long\def\ptnl{\alpha}
\global\long\def\currsp{\mathcal{I}}
\global\long\def\volt{V}
\global\long\def\intv{\mathbf{t}}
\global\long\def\intc{t}
\global\long\def\intsp{\mathcal{T}}
\global\long\def\frcv{\mathbf{f}}
\global\long\def\frcc{f}
\global\long\def\frcsp{\mathcal{F}}
\global\long\def\velv{\mathbf{V}}
\global\long\def\velc{V}
\global\long\def\disv{\mathbf{E}}
\global\long\def\disc{E}
\global\long\def\posn{\mathbf{x}}
\global\long\def\area{\mathbf{A}}
\global\long\def\relp{\mathbf{L}}
\global\long\def\chn{c}

\title[Electrodynamics and Stress Theory]{Generalized Electrodynamics as a Special Case of Metric Independent
Stress Theory}

\author{Reuven Segev}

\curraddr{Reuven Segev\\
Department of Mechanical Engineering\\
Ben-Gurion University of the Negev\\
Beer-Sheva, Israel\\
rsegev@bgu.ac.il}


\thanks{\today}

\begin{abstract}
We use a metric invariant stress theory of continuum mechanics to
formulate a simple generalization of the the basic variables of electrodynamics
and Maxwell's equations to general differentiable manifolds of any
dimension, thus viewing generalized electrodynamics as a special case
of continuum mechanics. The basic variable is the potential, or a
variation thereof, which is represented as an $r$-form in a $d$-dimensional
spacetime. The stress for the case of generalized electrodynamics,
is assumed to be represented by a $(d-r-1)$-form, a generalization
of the Maxwell $2$-form.
\end{abstract}

\maketitle

\section{Introduction}

Metric independent, or pre-metric, aspects of electrodynamics have
been studied since the beginning of the 20th century. Whittaker \cite[pp. 192--196]{Whittaker53},
attributes the first work in this direction to Kottler \cite{Kottler1922}
while Truesdell and Toupin \cite[Section F]{TruesdellToupin60} attribute
the main contribution to van Dantzig \cite{dantzig}. In recent decades,
renewed interest in the subject led to further work in which notions
of modern differential geometry have been utilized (see for example
\cite{Hehl2003,Kaiser2004,Hehl-Itin06}).

In \cite{Segev1986,Segev2002} we proposed a metric invariant formulation
of continuum mechanics where the major objective was to introduce
a metric invariant notion of stress. In particular, using the metric
invariant formulation of Maxwell's equations in spacetime, and a metric
invariant version of the Lorentz force, we discussed in \cite{Segev2002,SegevRevMMAS2012}
the stress energy momentum tensor of electrodynamics.adsf

In this note, we want to use the same setting for metric invariant
stress theory of continuum mechanics in order to formulate a simple
generalization of the the basic variables of electrodynamics and Maxwell's
equations to general differentiable manifolds of any dimension $d$.
Thus, we present generalized electrodynamics as a special case of
continuum mechanics. Here, the basic variable is the potential, or
a variation thereof, which is represented as an $r$-form in a $d$-dimensional
spacetime. The stress for the case of generalized electrodynamics,
is assumed to be represented by a differential $(d-r-1)$-form, a
generalization of the Maxwell $2$-form.

We recall that for classical continuum mechanics, one assumes that
the forces on a body $\reg\subset\reals^{3}$ are given in terms of
a vector field $\bfc$ defined in the physical space and a surface
force $\sfc_{\reg}$ defined on the boundary $\bdry\reg$ of $\reg$.
The virtual power of the forces on $\reg$ for a virtual velocity
field $\vf$ is given by
\begin{equation}
P_{\reg}=\int_{\reg}\bfc\cdot\vf\dV+\int_{\bdry\reg}\sfc_{\reg}\cdot\vf\dA.
\end{equation}
It is further recalled that if the dependence of $\sfc_{\reg}$ on
the body \textbf{$\reg$} satisfies Cauchy's postulates, then, there
is a $3\times3$ tensor field, the Cauchy tensor field $\st$ such
that $\sfc_{\reg}(x)=\st(x)(\nor(x))$ where $\nor(x)$ is the outwards
pointing normal to the boundary of $\reg$ at $x\in\bdry\reg$. Thus,
one has
\begin{equation}
P_{\reg}=\int_{\reg}\bfc\cdot\vf\dV+\int_{\bdry\reg}\st\trps(\vf)\cdot\nor\dA
\end{equation}
where the standard definition of the transpose has been used.

It is also noted that balance of moment of momentum implies traditionally
that the stress tensor $\st$ is symmetric. However, we want to examine
the case where $\st$ is skew symmetric. In this case, using the Levi-Civita
symbol, we may define a vector field $\maxw$ whose components are
given by
\begin{equation}
\maxw_{p}=\half\eps_{pjk}\st_{jk},
\end{equation}
such that $\st\trps(\vf)=\maxw\cross\vf$. Thus, assuming for simplicity
that $\bfc=0$, one has
\begin{equation}
P_{\reg}=\int_{\bdry\reg}(\maxw\cross\vf)\cdot\nor\dA.
\end{equation}
Using Gauss's theorem and the identity $\nabla\cdot(\maxw\cross\vf)=(\nabla\cross\maxw)\cdot\vf-\maxw\cdot(\nabla\cross\vf)$,
we have
\begin{equation}
P_{\reg}=\int_{\reg}(\nabla\cross\maxw)\cdot\vf\dV-\int_{\reg}\maxw\cdot(\nabla\cross\vf)\dV.
\end{equation}
Setting
\begin{equation}
\fflow=\nabla\cross\maxw,\qquad\frdy=\nabla\cross\vf,\label{eq:MaxSimp1}
\end{equation}
so that
\begin{equation}
\nabla\cdot\fflow=0,\qquad\nabla\cdot\frdy=0,\label{eq:MaxSimp2}
\end{equation}
the power may be written in the form
\begin{equation}
P_{\reg}=\int_{\reg}\fflow\cdot\vf\dV-\int_{\reg}\maxw\cdot\frdy\dV.
\end{equation}

We finally observe that in case we interpret $\vf$ as the vector
potential of magnetostatics, interpret $\maxw$ as the magnetic field
intensity, interpret $\frdy$ as the magnetic field and interpret
$\fflow$ as the current density, Equations (\ref{eq:MaxSimp1}\ref{eq:MaxSimp2})
are simply the restriction of Maxwell's equations to magnetostatics.

It is the generalization of this procedure to a general $d$-dimensional
manifold that we consider below.

\section{A Review of Metric Invariant Stress Theory}

We consider the following setting for metric independent continuum
mechanics using the Eulerian (spatial) point of view. A smooth $\dims$-dimensional
manifold $\man$ will denote either the standard space manifold of
continuum mechanics or space-time. Elements of a vector bundle $\vb:\vbts\to\man$
will represent values of virtual generalized velocities. Virtual generalized
velocities need not be of kinematic character and may involve generalized
coordinates, internal degrees of freedom, order parameters, etc.

For a generic vector bundle $\pi:V\to\man$, we will use the notation
$C^{\infty}(V)$ for the vector space of smooth sections. The notation
$\ext^{r}T^{*}\man$ will be used for the bundle of alternating $r$-tensors
over $\man$. For this case, $V=\ext^{r}T^{*}\man$ and $\forms^{r}(\man)=C^{\infty}\left(\ext^{r}T^{*}\man\right)$
is the space of differential $r$-forms.

\subsection{\label{sub:Force-Fields}Generalized Force Fields on Manifolds}

As is customary in continuum mechanics, forces on regions are characterized
as either body forces or surface forces. For general differentiable
manifolds, devoid of any Riemannian metric structure, forces are represented
by the power they produce for the various virtual velocities. Let
$\reg$ be a compact $d$-dimensional admissible region, a smooth
orientable submanifold or a chain, in $\man$ with boundary $\bdry\reg$
so that integration theory of differential forms on manifolds applies.
We will make no distinction in the notation between the vector bundle
$\vb$ and its restriction to $\reg$. A virtual velocity field over
$\reg$ will be represented by a section $\vf:\reg\to\vbts$. We let
$\L{\vbts,\ext^{p}T^{*}\reg}$ denote the vector bundle over $\reg$
whose fiber at $x\in\reg$ is the space of linear mappings $\vbts_{x}\to\ext^{p}T_{x}^{*}\reg$
and we use the analogous notation for $\man$.

A \emph{body force} is a section $\bfc:\man\to L(\vbts,\ext^{d}(T^{*}\man))$.
Given a section $\vf$ of $\vb$, the $d$-form ${\bf \bfc}(\vf)$,
given by $\bfc(\vf)(x)=\bfc(x)(\vf(x))$, represents the power density
and may be integrated over $\reg$.Thus, the total power expended
by the body force for a velocity field $\vf:\reg\to\vbts$ on $\reg$
is
\begin{equation}
\int_{\reg}\bfc(\vf).
\end{equation}
A \emph{boundary force} field, or a \emph{surface force }field, on
$\bdry\reg$ is a section
\begin{equation}
\sfc_{\reg}:\bdry\reg\to\L{W,\ext^{d-1}(T^{*}\bdry\reg)},
\end{equation}
where again, for the sake of simplicity, we make no distinction in
the notation between $\vbts$ and $\vbts\resto{\bdry\reg}$. Thus,
for any $(d-1)$-dimensional submanifold $D$ of $\bdry\reg$, the
power expended by the surface force for a velocity field $u$ defined
on $\bdry\reg$ is given by
\begin{equation}
\int_{D}\sfc_{\reg}(u).\label{eq:PowerOfSurfaceForces}
\end{equation}
The total power expended by both the body force and surface force
over the region $\reg$ and its boundary is viewed as the action of
a linear functional $F_{\reg}$ on the virtual velocity field $\vf$
and is given therefore by
\begin{equation}
\fc_{\reg}(\vf)=\int_{\reg}\bfc(\vf)+\int_{\bdry\reg}\sfc_{\reg}(\vf).\label{eq:TotalPowerOnARegion}
\end{equation}

\subsection{Smooth Stress Fields on Manifolds}

A \emph{traction stress} is a section $\st$ of $\L{\vbts,\ext^{d-1}T^{*}\man}$.
 Without changing the notation, we also view $\st$ as a linear mapping
\begin{equation}
\st:C^{\infty}(\vbts)\tto\forms^{d-1}(\man)
\end{equation}
so that for a section $\vf$, the $(d-1)$-form $\st(\vf)$, is given
$\st(\vf)(x)=\st(x)(\vf(x))$, Physically, $\st(\vf)$ is a flux field
representing a flux of power. Given an oriented $(d-1)$-submanifold
$D\subset\man$, the $(d-1)$-form $\st(\vf)$ may be restricted to
vectors tangent to $D$ using the inclusion $\incl_{D}:D\to\man$.
Let $T\incl_{D}:TD\to T\man$ denote the tangent to the inclusion,
we have a mapping, the pullback of differential forms,
\begin{equation}
\incl_{D}^{*}:\forms^{d-1}(\man)\tto\forms^{d-1}(D),
\end{equation}
with
\begin{equation}
\incl_{D}^{*}\go(x)(\lisub v,{d-1})=\go(x)(T_{x}\incl_{D}(v_{1}),\dots,T_{x}\incl_{D}(v_{d-1})),\qquad v_{1},\dots,v_{d-1}\in T_{x}D.
\end{equation}
In accordance with the notation introduced above, $C^{\infty}\left(\L{\vbts,\ext^{d-1}T^{*}\man}\right)$
is the space of smooth stress fields and $C^{\infty}\left(\L{\vbts,\ext^{d-1}T^{*}D}\right)$
is the space of smooth surface force fields on $D$. We thus have
a restriction mapping
\begin{equation}
\incl_{D}^{*}\comp\st:C^{\infty}\left(\L{\vbts,\ext^{d-1}T^{*}\man}\right)\tto C^{\infty}\left(\L{\vbts,\ext^{d-1}T^{*}D}\right).
\end{equation}

The total power expended over $D$ for the vector field $u:D\to\vbts\resto D$
is therefore given by
\begin{equation}
\int_{D}\incl_{D}^{*}(\st(u)).\label{eq:PowerFromTractionStress}
\end{equation}
Thus, a traction stress induces a surface on any orientable hypersurface
$D$ by the generalized Cauchy formula
\begin{equation}
\sfc=\incl_{D}^{*}\comp\st.
\end{equation}

Using Stokes's theorems, the power of the induced surface force $\sfc_{\reg}$
corresponding to a generalized velocity field $u$ on $\bdry\reg$
may be written now as
\begin{equation}
\begin{split}\int_{\bdry\reg}\sfc_{\reg}(u) & =\int_{\bdry\reg}\incl_{\bdry\reg}^{*}(\st(u)),\\
 & =\int_{\reg}\dee(\st(u)),
\end{split}
\end{equation}
so that
\begin{equation}
\fc_{\reg}(\vf)=\int_{\reg}\bfc_{\reg}(\vf)+\dee(\st(\vf)).\label{eq:TotalPowerWithTrackStesses}
\end{equation}

It is noted that the values of the $d$-form in the integral above
are linear in the jet extension $j^{1}(\vf)$ which as section of
the jet bundle $\Jet{\vbts}$. Thus, there is a section $\stm$ of
$\L{\Jet{\vbts},\ext^{d}(T^{*}\man)}$ such that $\bfc(\vf)+\dee(\st(\vf))=\stm(j(\vf))$.
We refer to $\stm$ as the \emph{variational stress}. This may be
summarized by
\begin{equation}
\fc_{\reg}(\vf)=\int_{\reg}\bfc(\vf)+\dee(\st(\vf))=\int_{\reg}\bfc(\vf)+\int_{\bdry\reg}\sfc_{\reg}(\vf)=\int_{\reg}\stm(j(\vf))\label{eq:PrincipleVirtualPower}
\end{equation}
which is the metric independent version of the principle of virtual
work. In contrast with classical setting for stress theory in continuum
mechanics, for the metric independent analysis one has to make a distinction
between the traction stress $\st$ that induces the surface traction
using the generalized Cauchy formula and the variational stress which
determines the density of power via the principle of virtual power.
We emphasize that in the general case, since the jet of a section
cannot be decomposed invariantly into the value of a section and the
value of the derivative of a section, the variational stress cannot
be decomposed uniquely into a component that is dual to the values
of the section and a component that is dual do the valued of the derivative.
For example, unlike classical continuum mechanics, one cannot require
that the component that is dual to the values of the section should
vanish.

\section{Application of Stress Theory to Electrodynamics}

In generalized electrodynamics, the foregoing analysis specializes
to the case where $\vbts$ is the vector bundle of $r$-alternating
tensors, $r\les d-1$, \ie, $\vbts=\ext^{r}T^{*}\man$. Thus, we
have the sections
\begin{equation}
\begin{split}\sfc_{\reg} & \in C^{\infty}\left(\L{\ext^{r}T^{*}\man,\ext^{d-1}T^{*}\bdry\reg}\right),\\
\bfc & \in C^{\infty}\left(\L{\ext^{r}T^{*}\man,\ext^{d}T^{*}\man}\right),\\
\st & \in C^{\infty}\left(\L{\ext^{r}T^{*}\man,\ext^{d-1}T^{*}\man}\right).
\end{split}
\end{equation}
The sections of $\vbts=\ext^{r}T^{*}\man$ are interpreted now as
variations of the potential forms (vector potential in the traditional
formulation) and a generic such variation will be denoted as $\ga$
(rather than $\vf$ as above). The manifold $\man$ is interpreted
as space-time and it is natural to assume here that $\bfc=0$ .

What characterizes generalized electrodynamics is the following assumption
which is of a constitutive nature: Each traction stress $\st$ may
be represented by a $(d-r-1)$-form $\maxw$, the \emph{generalized
Maxwell form}, as
\begin{equation}
\st(\ga)=\maxw\wedge\ga\in C^{\infty}\left(\ext^{d-1}T^{*}\man\right),
\end{equation}
for any section $\ga$ of $\ext^{r}T^{*}\man$. This assumption implies
that for each region $\reg\subset\man$,
\begin{equation}
\begin{split}\fc_{\reg}(\ga) & =\int_{\bdry\reg}\st(\ga),\\
 & =\int_{\bdry\reg}\maxw\wedge\ga,\\
 & =\int_{\reg}\dee(\maxw\wedge\ga),\\
 & =\int_{\reg}\dee\maxw\wedge\ga+(-1)^{d-r-1}\int_{\reg}\maxw\wedge\dee\ga.
\end{split}
\label{eq:EMforces}
\end{equation}

Using the terminology of de Rham currents, we denote the the $d$-current
\begin{equation}
\go\longmapsto\int_{\reg}\go
\end{equation}
simply by $\reg$. The boundary of this current is the $(d-1)$-current
$\bdry\reg$ which satisfies
\begin{equation}
\bdry\reg(\psi)=\reg(\dee\psi)
\end{equation}
for any $(d-1)$-form $\psi$. In addition, for a de Rham $r$-current
$T$ and a smooth $p$-form $\vph$, with $p\les r$, the $(r-p)$-current
$T\fcontr\vph$ is defined by
\begin{equation}
T\fcontr\vph(\go)=T(\vph\wedge\go).
\end{equation}
Thus, Equation (\ref{eq:EMforces}) may be rewritten as
\begin{equation}
\fc_{\reg}(\ga)=\bdry\reg(\maxw\wedge\ga)
\end{equation}
so that
\begin{equation}
F_{\reg}=\bdrr\fcontr\maxw.
\end{equation}

One may set
\begin{equation}
\frdy=\dee\ga,\qquad\fflow=\dee\maxw,
\end{equation}
$\frdy\in C^{\infty}\left(\ext^{r+1}T^{*}\man\right)$, $\fflow\in C^{\infty}(\ext^{d-r})$,
so that
\begin{equation}
\dee\frdy=0,\quad\text{and}\quad\dee\fflow=0.
\end{equation}
If one views $\frdy$ as a generalization of the Faraday form and
$\fflow$ as a generalization of the $4$-charge-current density of
electrodynamics, the equations above generalize Maxwell's equations.

The total virtual power is now represented by
\begin{equation}
\fc_{\reg}(\ga)=\int_{\reg}\fflow\wedge\ga+(-1)^{d-r-1}\maxw\wedge\frdy.
\end{equation}
We may also write
\begin{equation}
\begin{split}\stm(j^{1}(\ga)) & =\fflow\wedge\ga+(-1)^{d-r-1}\maxw\wedge\frdy,\\
 & =\fflow\wedge\ga+(-1)^{d-r-1}\maxw\wedge\dee\ga.
\end{split}
\end{equation}
It follows that in the case of generalized electrodynamics, the variational
stress may be decomposed invariantly in the form
\begin{equation}
\stm(j^{1}(\ga))=\stm_{0}(\ga)+\stm_{1}(\dee\ga),
\end{equation}
where,
\begin{equation}
\stm_{0}(\ga)=\fflow\wedge\ga,\qquad S_{1}(\dee\ga)=(-1)^{d-r-1}\maxw\wedge\dee\ga.
\end{equation}

For a $(d-r)$-form $\vph$, consider the de Rham $r$-current $\reg\fcontr\vph$
so that
\begin{equation}
\reg\fcontr\vph(\ga)=\int_{\reg}\vph\wedge\ga.
\end{equation}
Thus, we may write
\begin{equation}
\begin{split}\fc_{\reg}(\ga) & =\reg\fcontr\fflow(\ga)+(-1)^{d-r-1}\reg\fcontr\maxw(\dee\ga),\\
 & =(\reg\fcontr\fflow+(-1)^{d-r-1}\bdry(\reg\fcontr\maxw))(\ga),
\end{split}
\end{equation}
so that
\begin{equation}
\fc_{\reg}=\reg\fcontr\fflow+(-1)^{d-r-1}\bdry(\reg\fcontr\maxw).
\end{equation}

\newpage{}

\bigskip{}

\noindent \textbf{\textit{Acknowledgments.}} This work was partially
supported by the Pearlstone Center for Aeronautical Engineering Studies
at Ben-Gurion University.

\end{document}